# Risk-Aware Objective-Based Forecasting in Inertia Management

Haipeng Zhang, Ran Li, *Member, IEEE*, Yan Chen, *Student Member, IEEE*, Zhongda Chu, *Student Member, IEEE*, Mingyang Sun, *Member, IEEE*, and Fei Teng, *Senior Member, IEEE*

*¹ Abstract*--The objective-based forecasting considers the asymmetric and non-linear impacts of forecasting errors on decision objectives, thus improving the effectiveness of its downstream decision-making process. However, existing objective-based forecasting methods are risk-neutral and not suitable for tasks like power system inertia management and unit commitment, of which decision-makers are usually biased toward risk aversion in practice. To tackle this problem, this paper proposes a generic risk-aware objective-based forecasting method. It enables decision-makers to customize their forecasting with different risk preferences. The equivalence between the proposed method and optimization under uncertainty (stochastic/robust optimization) is established for the first time. Case studies are carried out on a Great Britain 2030 power system with system operational data from National Grid. The results show that the proposed model with deterministic optimization can approximate the performance of stochastic programming or robust optimization at only a fraction of their computational cost.

## Nomenclature

*Sets and Indexes*:

| | |
|---|---|
| $I / i$ | Set/index of units. |
| $S / s$ | Set/index of training scenarios. |
| $N / n$ | Set/index of testing scenarios. |

*Variables*:

| | |
|---|---|
| $R_i^D / R_i^B$ | Day-ahead/real-time regulating reserve schedule of unit $i$. |
| $R_{is}^D / R_{is}^B$ | Day-ahead/real-time regulating reserve schedule of unit $i$ for scenarios $s$. |
| $\theta$ | Parameter vector for the forecasting model. |
| $VaR$ | Value at risk. |

*Constants*:

| | |
|---|---|
| $\hat{H}_s$ | Forecasted value of inertia for scenarios $s$. |
| $x_s$ | The input feature vector for scenarios $s$. |
| $\Delta f / \Delta f_{lim}$ | Frequency deviation/Maximum permissible frequency deviation. |
| $\Delta f^{ss} / \Delta f_{lim}^{ss}$ | Steady-state frequency deviation/Maximum steady-state frequency deviation. |
| $\Delta \dot{f} / \Delta \dot{f}_{lim}$ | RoCoF/Maximum permissible RoCoF. |
| $H$ | System inertia. |
| $R$ | Regulating reserve. |
| $T_d$ | Fully delivered time of primary frequency response. |
| $\Delta P_L$ | Loss of generation. |
| $D$ | Load-dependent damping in the system. |
| $C_i^D / C_i^B$ | Day-ahead/real-time regulating reserve bid price for unit $i$. |
| $R_i^{max}$ | Regulating reserve capacity limit available for unit $i$. |
| $\hat{R}_{min,s}$ | Forecasted value of minimum regulating reserve for scenarios $s$. |
| $\alpha$ | Weighting parameter for objective versus risk tradeoff. |
| $\beta$ | Confidence level. |

*Operators*:

| | |
|---|---|
| $|\cdot|$ | Cardinality of a set. |
| $\mathbb{E}(\cdot)$ | Expectation operator. |

## I. Introduction

The key role of forecasting in power systems is to guide optimization models to make effective decisions. In current practice, forecasting and optimization are independent because their objectives are misaligned. Existing forecasting methods aim to minimize statistical error metrics, such as mean square error (MSE) and mean absolute percentage error (MAPE), while optimization models aim to minimize/maximize decision objectives, such as system operating cost and reliability.

An underlying assumption here is that a statistically accurate forecast will guarantee effective decision-making. However, recent studies have demonstrated that the impact of forecasting

This work was supported by the National Natural Science Foundation of China under Grant 52107114. (*Corresponding author: Ran Li*)

Haipeng Zhang and Ran Li are with the Key Laboratory of Control of Power Transmission and Conversion, Ministry of Education, and Shanghai Non-Carbon Energy Conversion and Utilization Institute, Shanghai Jiao Tong University, Shanghai 200240, China (e-mail: 13122270150@163.com; rl272@sjtu.edu.cn).

Yan Chen and Mingyang Sun are with the State Key Laboratory of Industrial Control Technology and the College of Control Science and Engineering, Zhejiang University, Hangzhou 310027, China (e-mail: chenyan16@zju.edu.cn; mingyangsun@zju.edu.cn).

Zhongda Chu and Fei Teng are with the Department of Electrical and Electronic Engineering, Imperial College London, SW7 2AZ London, U.K. (e-mail: zc4915@ic.ac.uk; f.teng@imperial.ac.uk).

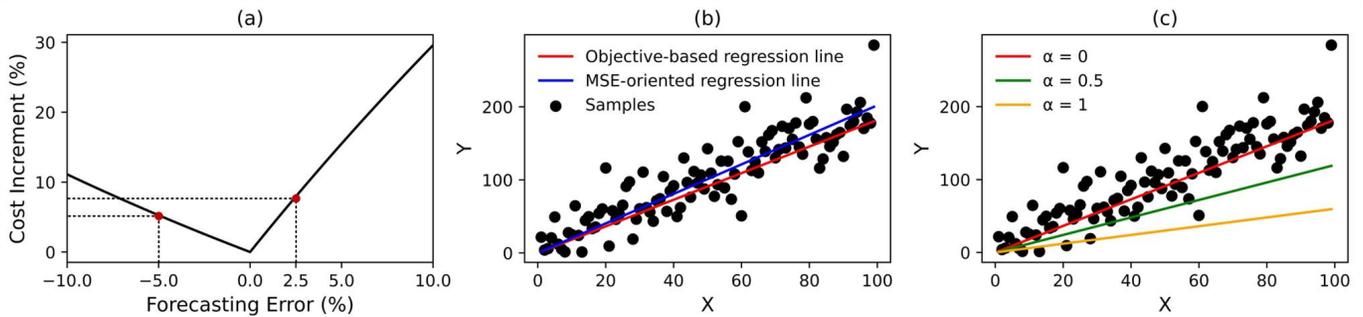

Fig. 1. (a) Asymmetric impact of forecasting error on cost increment. (b) Difference between objective-based and MSE-oriented regression line. (c) $\alpha$ represents the degree of risk aversion, and a higher value indicates a higher aversion to risk. The regression line when $\alpha = 0$ is equal to the objective-based one.

errors on decision objectives could be asymmetric, non-monotonic, and non-linear, leading to a mismatch between more accurate forecasts and more effective decision-making [1]. Fig. 1(a) depicts a simple scenario to illustrate the asymmetric impact of positive or negative forecasting errors on the system operating cost. The cost increment caused by positive errors is higher than that caused by negative errors. Comparing two forecasts with the forecast errors -5% and 2.5%, the latter (2.5%) is statistically more accurate but operationally more costly.

To align forecasting performance with decision objectives, a common approach is to deploy probabilistic forecasting followed by uncertainty optimization (PF-UO). Probabilistic forecasting techniques create scenarios according to probability density functions or quantiles. Uncertainty optimization will consider the objectives under some or all scenarios to make decisions based on risk preferences. If decision-makers are risk-neutral, they could resort to stochastic programming (SP), which smooths out the decision objectives of each singular scenario by optimizing the total expectation [2], [3]. If they are risk-averse, then they could opt for robust optimization (RO) or risk measures such as conditional value at risk (CVaR) and regret, representing a tradeoff between decision effectiveness and risk mitigation [4].

Although different PF-UO approaches have been proposed in the literature, most system operators in practice still adhere to the approach of point forecasting followed by deterministic optimization, being a simpler, more transparent, and hence easier to be accepted approach by different stakeholders [5]. The other reason is that uncertain optimization tends to be more computationally insensitive than deterministic optimization [6].

In order to improve the effectiveness of deterministic optimization, the objective-based forecasting (OBF) methods were proposed and have been widely applied in the energy sector as summarized in Table I. Reference [7], [13] replace traditional statistical error metrics with cost-oriented loss functions constructed through a piecewise linearization method to guide more economical load and wind generation forecasting. In [8], two machine learning methods are used to integrate forecasting and decision-making. A method for improving the forecasting of renewable power production based on extra information and specific use is proposed in [9]. In [10], the linear wind generation forecasting model is integrated into the unit commitment model. In [11], a cost-oriented probabilistic forecasting method for wind power was proposed. Reference [12] proposed a mixed-integer model to find the value of net demand that must be cleared in the forward market to minimize the system operating cost. In [14], [15], cost-oriented loss and mean square error loss are combined for optimization, improving the decision-making while ensuring the required accuracy. References [16]-[18] calculate the reserve requirements based on stochastic bilevel programming, which enhances the inter-temporal coordination of energy and reserve markets. Reference [19] on the basis of [10], integrates the economic dispatch model to boost the market economy further. The key innovation of OBF was to train point forecasting models with the objective of minimizing the expectation of decision objectives instead of statistical error metrics. An intuitive explanation of OBF is shown in Fig. 1(b). The blue line represents the MSE-oriented forecasting which is essentially a 50% quantile regression. The red line represents the objective-based forecast, which in this case is close to the 40% quantile regression, reflecting that the decision effectiveness of over-forecasting is better than that of under-forecasting.

TABLE I
REVIEW OF APPLICATIONS OF OBJECTIVE-BASED FORECASTING

| Reference | Forecast | Objective |
|---|---|---|
| [7]-[11] | Renewable energy | Benefits for renewable energy producers |
| [12]-[14] | Load | System operating costs |
| [15] | Electricity price | Net energy benefits |
| [16-17] | Reserve | System operating costs |
| [18] | Load and reserve | Reserves allocation costs plus energy dispatch costs |
| [19] | Renewable energy and reserve | System operating costs |

Despite the promising advances in OBF methods, two research gaps still exist, as shown in Fig. 2. Research gap 1 is that existing objective-based forecasting methods are risk-neutral and unsuitable for tasks with risk preferences. For example, power system operators tend to be risk-averse [3], [20], especially regarding security issues. For inertia forecasting, its downstream scheduling decisions based on frequency security constraints are biased towards conservative to further reduce the risk of frequency violation and power blackouts. To the best of the authors' knowledge, there is no prior study exists which considers risk preference in OBF.

To bridge this research gap, a risk-aware objective-based forecasting (RAOBF) method is proposed in this paper. The proposed method aims to expand the boundary of the OBF method by including risk preferences. An example is shown in Fig. 1(c). The parameter $\alpha$ controls the degree of risk aversion. The red line represents the forecasting with $\alpha = 0$, equivalent to traditional OBF. The green and orange lines represent the OBF with different risk preferences ($\alpha = 0.5$ and $\alpha = 1$, respectively). The larger the $\alpha$, the more conservative the forecasting and the less risk in system operation.

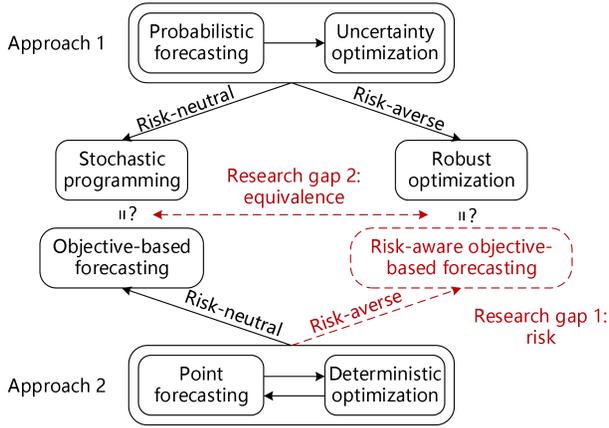

Fig. 2. Two approaches to aligning forecasting performance with decision objectives

Research gap 2 is whether the proposed RAOBF could approximate the decision of PF-UO with different risk preferences. This paper investigates the equivalence of RAOBF and PF-UO with error analysis. For risk-neural cases, the error mainly comes from the difference between the scenario set of SP and the training scenario set of RAOBF. For risk-aversion cases, the error source is the gap between the uncertainty set of RO (equivalent to VAR) and the risk parameters of RAOBF (equivalent to CVAR).

To this end, the key contributions of this paper are as follows:

1) A generic RAOBF method is proposed. The proposed method consists of an objective part aimed at minimizing the expectation of decision objectives and a risk part aimed at reducing the tail risk. Risk parameters can be adjusted to balance the tradeoff between the objective and risk parts to obtain forecasts with different degrees of risk aversion.

2) The equivalence between RAOBF and PF-UO is analyzed. With similar scenario sets and matched parameters, RAOBF can approximate the performance of PF-UO at only a fraction of their computational cost.

3) The proposed method is applied in inertia management. Through case studies, we demonstrate the effectiveness of the proposed method and analyze the advantages of the proposed method in terms of cost-effectiveness and risk mitigation compared with traditional point forecasting methods.

The rest of the paper is organized as follows: Section II presents the RAOBF method. Section III analyzes the equivalence between RAOBF and PF-UO. Section IV describes the application of the proposed method in inertia management. Section V verifies the proposed method, and Section VI makes the conclusions of the work.

## II. RISK-AWARE OBJECTIVE-BASED FORECASTING METHOD

In this section, we first describe the two-stage decision-making process as an application scenario of RAOBF. Then, the generic form of the OBF method is introduced. Finally, we propose a RAOBF method on this basis.

### A. Two-Stage Decision Process

Without loss of generality, we consider a two-stage decision process. Stage I takes the forecasted value $\hat{u}$ as input and provides the decision $\Phi^I$ that minimizes the objective $C^I$, which could be done in the day-ahead stage. The decision process of stage I can be expressed as below:

$$\min_{\Phi^I} \quad C^I(\Phi^I) \quad (1.1)$$

$$s.t. \quad F(\Phi^I, \hat{u}) \quad (1.2)$$

where $F$ represents the constraint set of stage I. Stage II takes the true value $u$ and the decision $\Phi^I$ from stage I as input, providing the decision $\Phi^{II}$ that minimizes the objective $C^{II}$. The decision process of stage II can be expressed as below:

$$\min_{\Phi^{II}} \quad C^{II}(\Phi^{II}) \quad (2.1)$$

$$s.t. \quad G(\Phi^I, \Phi^{II}, u) \quad (2.2)$$

where $G$ represents the constraint set of stage II.

### B. Objective-Based Forecasting

In the above decision process, the total objective, i.e., the sum of $C^I$ and $C^{II}$, is determined by $\hat{u}$ and $u$, which can be regarded as a function of $\hat{u}$ and $u$, as below:

$$C(\hat{u}, u) = \min_{\Phi^I, \Phi^{II}} \quad C^I(\Phi^I) + C^{II}(\Phi^{II}) \quad (3.1)$$

$$s.t. \quad F(\Phi^I, \hat{u}) \quad (3.2)$$

$$G(\Phi^I, \Phi^{II}, u) \quad (3.3)$$

Based on (3), we can train point forecasting models with the objective of minimizing the expectation of total objectives instead of statistical error metrics. This method of training point forecasting models can be defined as the OBF method, and its generic form is as below:

$$\min_{\boldsymbol{\theta}} \quad \frac{1}{|S|} \sum_{s \in S} C(\hat{u}_s, u_s) \quad (4.1)$$

$$s.t. \quad \hat{u}_s = f(\boldsymbol{\theta}, \boldsymbol{x}_s), \forall s \in S \quad (4.2)$$

where $f(\cdot)$ is defined as a function of the parameter vector $\boldsymbol{\theta}$ and the input feature vector $\boldsymbol{x}_s$, which can be linear or nonlinear, depending on the selected forecasting model (e.g., linear regression and neural network).

All the references [7]-[19] demonstrate that the OBF method can significantly improve the decision effectiveness of decision-making, although the statistical error may be slightly increased. However, the existing approach can not explicitly consider risk preferences in the decision-making process.

### C. Risk-Aware Objective-Based Forecasting

To incorporate risk mitigation, the RAOBF method can be developed by combining the risk measure on the basis of the OBF method. The proposed method is to train forecasting models with the objective of minimizing the expectation of total objectives, including the risk of objective variability (measured in terms of CVaR), which can be defined as below:

$$\min_{\Xi} \frac{(1-\alpha)}{|S|} \sum_{s \in S} C(\hat{u}_s, u_s) + \alpha \text{CVaR} \quad (5.1)$$

$$\text{s.t.} \quad \hat{u}_s = f(\boldsymbol{\theta}, \boldsymbol{x}_s), \forall s \in S \quad (5.2)$$

$$\text{CVaR} = \text{VaR} + \frac{1}{1-\beta} \sum_{s \in S} \frac{\mu_s}{|S|} \quad (5.3)$$

$$C(\hat{u}_s, u_s) - \mu_s - \text{VaR} \leq 0, \forall s \in S \quad (5.4)$$

$$\mu_s \geq 0, \forall s \in S \quad (5.5)$$

where $\Xi = \{\boldsymbol{\theta}, \text{VaR}, \mu_s, \forall s\}$ is the set of optimization variables. We divide the above optimization model into two parts:

*1) Objective part:* The objective part includes the first term of the objective function (5.1) and constraints (5.2). This part is similar to (4).

*2) Risk part:* The risk part includes the second term of the objective function (5.1) and constraints (5.3)-(5.5). The parameter $\alpha \in [0,1]$ represents the degree of risk aversion. The larger its value, the higher impact of the risk part in the model.

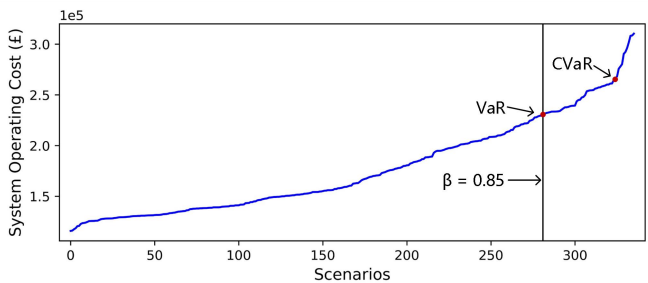

Fig. 3. Schematic diagram of VaR and CVaR.

This part aims to limit the volatility of total objectives by minimizing CVaR, thus introducing risk mitigation. CVaR measure, also known as tail VaR, refers to the expectation of total objectives exceeding VaR at a given confidence level $\beta$ and is an appropriate method to incorporate the risk management problem into the OBF method. In this paper, for a given $\beta$ (0, 1), VaR is defined as the maximum total objective in the training scenario set, which ensures that the probability of obtaining a total objective is less than this value in the training scenario set is greater than $\beta$. As $\beta$ increases, the gap between CVaR and VaR will gradually narrow. When $\beta$ = 1-, CVaR is approximately equal to VaR. An example is shown in Fig. 3. Given the confidence level of 0.85, the VaR refers to the system operating cost that is higher than 85% of scenarios, while the CVaR refers to the expectation of system operating costs exceeding the VaR. As discussed in [4], the VaR measure has inherent flaws:

a) The distribution of VaR is not continuous, which may lead to failures in optimization problems.

b) VaR cannot measure tail risk exceeding quantile $\beta$.

c) The premise of the VaR application is that the total objective is subject to the normal distribution, which is difficult to satisfy in practice.

Compared with VaR, CVaR can measure tail risk and does not need to be subject to normal distribution. In addition, the mathematical formulation of CVaR is convex and can be effectively controlled by linear programming. The mathematical formulation of CVaR can be expressed as the constraint (5.3). The constraints (5.4) and (5.5) ensure that the total objective for scenario $s$ maintains a non-negative difference with respect to the VaR, the value of which is $\mu_s$.

## III. Equivalence between the Proposed Method and PF-UO

The advantage of SP lies in cost savings compared to deterministic models [6]. The idea of SP is to smooth out every possible scenario by minimizing the expectation of total objectives. Possible scenarios are extrapolated from the training scenario set $S$ through the probabilistic forecasting model $g(\cdot)$, and all possible scenarios can be included in a scenario set $\mathcal{H}$:

$$g(S) \mapsto \mathcal{H} \quad (6)$$

In a two-stage decision-making process, there exists a forecasted value $\hat{u}_{sto}$ that makes the decision $\Phi^1$ of the deterministic optimization model of stage I equivalent to that of its SP-based corresponding model. The forecasted value $\hat{u}_{sto}$ can be obtained by solving the SP-based corresponding model, as below:

$$\hat{u}_{sto} = \arg\min_{\hat{u}} \mathbb{E}_{h \in \mathcal{H}}[C(\hat{u}, \tilde{u}_h)] \quad (7)$$

where $\tilde{u}_h$ represents the realization of the parameter $u$ in scenario $h$. The goal of the RAOBF ($\alpha = 0$) method is to enable a forecasting model to forecast $\hat{u}_{sto}$ when $\boldsymbol{x}$ is given, expressed as (8). In this way, in stage I, there is no need to deploy the SP-based corresponding model with high computational cost, only the deterministic optimization with low computational cost.

$$\hat{u}_{sto} = f(\boldsymbol{\theta}_{sto}, \boldsymbol{x}) + \varepsilon \quad (8)$$

where $\varepsilon$ represents the forecasting error. When $\alpha = 0$, only the objective part works. What the RAOBF method does is to estimate the parameter vector $\boldsymbol{\theta}_{sto}$ in (8) from the training scenario set $S$:

$$\boldsymbol{\theta}_{sto} \approx \arg\min_{\boldsymbol{\theta}} \mathbb{E}_{s \in S}[C(f(\boldsymbol{\theta}, \boldsymbol{x}_s), u_s)] \quad (9)$$

The gap between RACOF and PF-SP also depends on the difference between the training scenario set $S$ and the scenario set $\mathcal{H}$, the former contains similar scenarios, and the latter includes all possible scenarios. In Section V, we quantitatively analyzed that using $S$ to solve (5) enables RAOBF ($\alpha = 0$) to approximate the performance of PF-SP using $\mathcal{H}$ to a large extent.

The advantage of RO lies in risk mitigation compared with SP-based models [6]. The idea of RO is to minimize the total objective of the worst scenario in the uncertain set $\mathcal{U}$. The uncertain set $\mathcal{U}$ is usually represented by the variation range of the forecasted value $\hat{u}$, which is also obtained from the training scenario set $S$ through probabilistic forecasting model $g(\cdot)$:

$$g(S) \mapsto \mathcal{U} \tag{10}$$

Similarly, in a two-stage decision-making process, there exists a forecasted value $\hat{u}_{ro}$ that makes the decision $\Phi^I$ of the deterministic optimization model in stage I equivalent to that of its RO-based corresponding model. The parameter $\hat{u}_{ro}$ can be obtained by solving the RO-based corresponding model, as below:

$$\hat{u}_{ro} = \arg\min_{\hat{u}} \max_{\tilde{u} \in \mathcal{U}} [C(\hat{u}, \tilde{u})] \tag{11}$$

When $\alpha = 1$, the objective of the RAOBF method is only to minimize the CVaR measure. When $\beta = 1^-$, the CVaR is approximately minimizing the maximum VaR, i.e., the worst scenario. Therefore, the goal of the RAOBF ($\alpha = 1$, $\beta = 1^-$) method is to enable a forecasting model to forecast $\hat{u}_{ro}$ when $x$ is given. The forecasting model can be expressed as below:

$$\hat{u}_{ro} = f(\boldsymbol{\theta}_{ro}, \boldsymbol{x}) + \varepsilon \tag{12}$$

When $\alpha = 1$ and $\beta = 1^-$, what the RBCOF method does is to estimate the parameter vector $\boldsymbol{\theta}_{ro}$ in (12) from the training scenario set $S$:

$$\boldsymbol{\theta}_{ro} \approx \arg\min_{\boldsymbol{\theta}} \max_{s} [C(f(\boldsymbol{\theta}, \boldsymbol{x}_s), u_s)] \tag{13}$$

In addition to forecasting error, the gap between RACOF and PF-RO depends on the difference between CVaR and VaR. When $\beta \neq 1^-$, there is a difference between CVaR and VaR because CVaR considers the tail risk exceeding VaR. This difference leads to a gap between RAOBF and PF-RO, and the magnitude of the gap depends on the value of $\beta$ and the shape of the tail of the total objective distribution, which is quantitatively analyzed in Section V.

## IV. Application in Inertia Management

In this section, we first describe the process of frequency response service market clearing, then provide mathematical formulations for each component, and finally, we combine each component to build a risk-aware objective-based inertia forecasting model.

### A. Frequency Response Service Market

The frequency response service market is to be cleared through independent and sequential auctions, which are implemented in, for example, the Nordic area and Great Britain [21], [22]. The simplified decision-making process is illustrated in Fig. 4.

First, the system operator performs day-ahead inertia forecasting based on the input feature vector $x$ to obtain the forecasted inertia value $\hat{H}$. The input feature vector $x$ includes the day-ahead forecast of synchronous generation, load, renewable generation, etc. Then, the regulating reserve $\hat{R}$ is calculated through frequency security constraints, such as steady-state frequency deviation and frequency nadir. In the day-ahead market, the market operator provides optimal schedule $R^{D*}$ based on price-quantity offers submitted by flexible producers to minimize day-ahead cost $C^D$, balancing the regulating reserve requirement. Finally, the total regulating reserve requirement $R$ is realized close to delivery time. The market operator provides optimal schedule $R^{B*}$ that minimizes the balancing cost $C^R$ to cover the real-time regulating reserves' shortages. In this process, the forecasted inertia value determines the regulating reserve, which directly impacts the day-ahead costs and, in turn, the balancing costs.

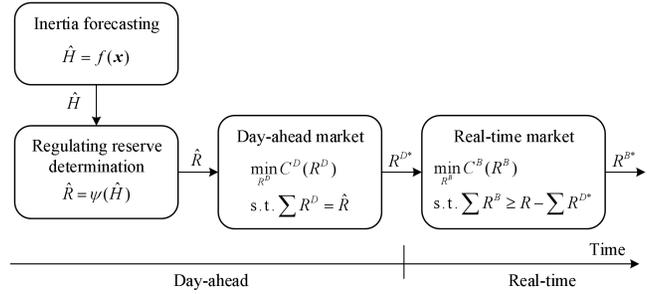

Fig. 4. Inertia management process.

### B. Mathematical Formulations for Each Component

*1) Inertia forecasting:* For simplicity, we consider a linear formulation as below:

$$\hat{H}_s = \boldsymbol{\theta}^T \boldsymbol{x}_s \tag{14}$$

where the input feature vector $\boldsymbol{x}_s$ includes the synchronous power generation, load, renewable power generation, the power flow with surrounding region systems, and the surrounding environmental state information.

*2) Regulating reserve determination:* Frequency security constraints include the frequency nadir constraint, the frequency steady-state deviation constraint, and the rate of change of frequency (RoCoF) constraint [23]. The mathematical formulations of the three are as below:

$$\Delta f = \frac{\Delta P_L^2 T_d}{4HR} \leq \Delta f_{\lim} \tag{15}$$

$$\Delta f^{ss} = \frac{\Delta P_L - R}{DP_D} \leq \Delta f_{\lim}^{ss} \tag{16}$$

$$\Delta \dot{f} = \frac{\Delta P_L}{2H} \leq \Delta \dot{f}_{\lim} \tag{17}$$

Inequation (15) shows that the frequency nadir depends on the inertia and regulating reserve, which are inversely proportional. Inequations (16) and (17) show that steady-state frequency deviation and RoCoF rely on the regulating reserve and inertia, respectively. Thus, the minimum regulating reserve $R_{\min}$ is a function of the maximum allowable frequency nadir and steady-state frequency deviation, which is equal to the greater value of regulating reserves determined by the frequency nadir constraint and the frequency steady-state deviation constraint. $R_{\min}$ can be computed as below:

$$R_{\min} = \max\left\{ \frac{\Delta P_L^2 T_d}{4H \Delta f_{\lim}}, \Delta P_L - DP_D \Delta f_{\lim}^{ss} \right\} \tag{18}$$

*3) Day-ahead market:* When $R_{\min}$ is determined, the day-ahead scheduling decisions can be obtained from the following deterministic optimization model:

$$\min_{R_i^D} \sum_{i \in I} C_i^D R_i^D \tag{19.1}$$

$$\text{s.t.} \quad 0 \leq R_i^D \leq R_i^{\max}, \forall i \in I \tag{19.2}$$

$$\sum_{i \in I} R_i^D = R_{\min} \tag{19.3}$$

The model objective (19.1) is to minimize day-ahead costs. Constraints (19.2) account for the quantity offers of each unit. Constraint (19.3) ensures the balance of the regulating reserve requirement.

*4) Real-time market:* Given the day-ahead scheduling decisions $R_i^{D*}$ and the total regulating reserve requirement $R$, the real-time scheduling decisions can be obtained from the following deterministic optimization model:

$$\min_{R_i^B} \sum_{i \in I} C_i^B R_i^B \tag{20.1}$$

$$\text{s.t.} \quad 0 \leq R_i^B \leq R_i^{\max} - R_i^{D*}, \forall i \in I \tag{20.2}$$

$$\sum_{i \in I} R_i^B \geq R - \sum_{i \in I} R_i^{D*} \tag{20.3}$$

The model objective (20.1) is to minimize real-time cost. Constraints (20.2) account for the real-time quantity offers of each unit. Constraint (20.3) ensures the real-time balance of the regulating reserve requirement.

*C. Risk-Aware Objective-Based Inertia Forecasting*

A risk-aware objective-based inertia forecasting model can be defined as the following model:

$$\min_{\Xi} \frac{(1-\alpha)}{|S|} \left[ \sum_{s \in S} \sum_{i \in I} (C_i^D R_{is}^D + C_i^B R_{is}^B) \right] + \alpha \text{CVaR} \tag{21.1}$$

$$\text{s.t.} \quad \hat{H}_s = \boldsymbol{\theta}^T \boldsymbol{x}_s, \forall s \in S \tag{21.2}$$

$$\hat{R}_{\min,s} = \frac{\Delta P_L^2 T_d}{4 \hat{H}_s \Delta f_{\lim}}, \forall s \in S \tag{21.3}$$

$$0 \leq R_{is}^D \leq R_i^{\max}, \forall i \in I, \forall s \in S \tag{21.4}$$

$$0 \leq R_{is}^B \leq R_i^{\max} - R_{is}^D, \forall i \in I, \forall s \in S \tag{21.5}$$

$$\sum_{i \in I} R_{is}^D = \hat{R}_{\min,s}, \forall s \in S \tag{21.6}$$

$$\sum_{i \in I} R_{is}^B \geq R_s - \sum_{i \in I} R_{is}^D, \forall s \in S \tag{21.7}$$

$$\text{CVaR} = \text{VaR} + \frac{1}{1-\beta} \sum_{s \in S} \frac{1}{|S|} \mu_s \tag{21.8}$$

$$\sum_{i \in I} (C_i^D R_{is}^D + C_i^B R_{is}^B) - \mu_s - \text{VaR} \geq 0, \forall s \in S \tag{21.9}$$

$$\mu_s \geq 0, \forall s \in S \tag{21.10}$$

where $\Xi = \{R_{is}^D, R_{is}^B, \forall i, \forall s; \mu_s, \forall s; \boldsymbol{\theta}, VaR\}$ is the set of optimization variables. Note that the frequency steady-state deviation constraint is removed since it is unrelated to inertia. Since the model contains quadratic constraints, it can be regarded as a quadratic constraint programming problem. Existing commercial solvers, such as Gurobi and CPLEX, can efficiently solve this model.

TABLE II
SELECTED FEATURES FOR RISK-AWARE OBJECTIVE-BASED INERTIA FORECASTING MODEL

| Feature type | Feature name |
| --- | --- |
| Interconnector flows | French, Dutch, Irish, East-West |
| Day-ahead forecasts | load, coal, solar, onshore/offshore wind power |
| Environment information | weekday or weekend, temperature |
| Inertia (Target) | total system inertia |

V. CASE STUDY

*A. Case Description*

*1) Testing system:* All the cases are conducted on a Great Britain 2030 power system [23] and solved by Gurobi 9.5.2 on a PC with the Intel i7-8650U CPU @ 1.90 GHz, 8GB RAM. The system includes 100 CCGTs (slower units with lower cost) each with a regulating reserve capacity of 50MW and 30 OCGTs (faster units with higher cost) each with a regulating reserve capacity of 20MW. CCGT is inflexible in real-time and OCGT is fully flexible in real-time. The reserve costs for CCGT and OCGT are £47/MW and £200/MW, respectively. The other system parameters are set: frequency response delivery time $T_d = 10$s, $\Delta f_{\lim} = 0.8$ Hz, and maximum power loss $\Delta P_L = 1800$ MW.

*2) Dataset:* We verify the performance of the proposed method on a simulated power system operational dataset of Great Britain from 2028 to 2030. The simulated dataset is extrapolated based on the power system historical data of Great Britain from 2016 to 2018 [24]. The simulated data includes various features such as system inertia, load, interconnector flows with outside, the short-term forecasted value of unit power, external environmental information, etc. The time granularity of these simulated data is 30 minutes. Since feature selection is not the main focus of this paper, we select some commonly used and effective features to form a feature set, summarized in Table II. The simulated data from 2030/1/1 to 2030/1/7 are used to test the performance of the proposed model. The training scenario set is built from the simulated data from 2028 to 2029 based on the method in [18].

*B. Evaluation Criteria*

Four different evaluation criteria are used to evaluate the performance of the proposed method. The first three are used to evaluate the economic efficiencies caused by forecasts, and the last one is used to evaluate the statistical errors of forecasts.

*1) Mean system operating cost (MSOC):* The MSOC is the mean system operating cost for all testing scenarios, calculated by solving the day-ahead market model (19) and the real-time market model (20), as below:

$$\text{MSOC} = \frac{1}{|N|} \sum_{n \in N} \text{SOC}_n \tag{22.1}$$

$$\text{SOC}_n = \sum_{i \in I} (C_i^D R_{in}^{D*} + C_i^B R_{in}^{B*}) \tag{22.2}$$

*2) Mean conditional value at risk (MCVaR):* The MCVaR is used to evaluate the average system operating risk for all

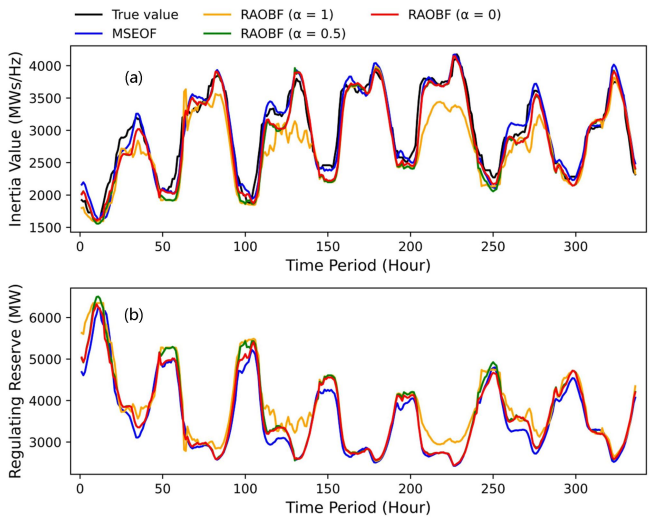

Fig. 5. Inertia forecasts and corresponding regulating reserves under different forecasting models.

TABLE III
COMPARATIVE ANALYSIS WITH DIFFERENT FORECASTING

|  | MSOC (£) | MCVaR (£) | MMaxSOC (£) | MAPE (%) |
|---|---|---|---|---|
| MSEOF | 181637.3 | 211962.1 | 218566.8 | **4.13** |
| RAOBF ($\alpha$ = 0) | **178348.4** | 195235.0 | 201839.7 | 4.84 |
| RAOBF ($\alpha$ = 0.5) | 179391.3 | 192473.4 | 199078.1 | 5.69 |
| RAOBF ($\alpha$ = 1) | 184032.5 | **191874.0** | **194165.2** | 9.68 |

testing scenarios. For each testing scenario, we first fix the optimal solution of the optimization model and then generate a random scenario set $\mathcal{K}$ to estimate the system operating cost incurred by the given solution under different random scenarios. Finally, we calculate the CVaR for this testing scenario. The MCVaR is the average CVaR for testing scenarios, which can be calculated as below:

$$\text{MCVaR} = \frac{1}{|N|}\sum_{n \in N} \text{CVaR}_n \quad (23.1)$$

$$\text{CVaR}_n = \mathbb{E}[\text{SOC} \,|\, \text{SOC} \geq \text{VaR}_n] \quad (23.2)$$

$$\text{VaR}_n = \{z \,|\, \text{Prob}\{\text{SOC} \geq z\} \leq 1 - \beta\} \quad (23.3)$$

*3) Mean maximum system operation cost (MMaxSOC):* The MMaxSOC is used to evaluate the average possible maximum system operating cost for all testing scenarios. For each testing scenario, we first fix the optimal solution of the optimization model and then generate a random scenario set $\mathcal{K}$ to estimate the system operating cost incurred by the given solution under different random scenarios. Finally, we calculate the possible maximum system operating cost (MaxSOC) for this testing scenario. The MMaxSOC is the average MaxSOC for testing scenarios, which can be calculated as below:

$$\text{MMaxSOC} = \frac{1}{|N|}\sum_{n \in N} \text{MaxSOC}_n \quad (24.1)$$

$$\text{MaxSOC}_n = \max\{\text{SOC}_1,...,\text{SOC}_{|\mathcal{K}|}\} \quad (24.2)$$

*4) Mean absolute percentage error (MAPE):* The MAPE is used to evaluate the error percentage between the forecasted value and the true value, which can be calculated as below:

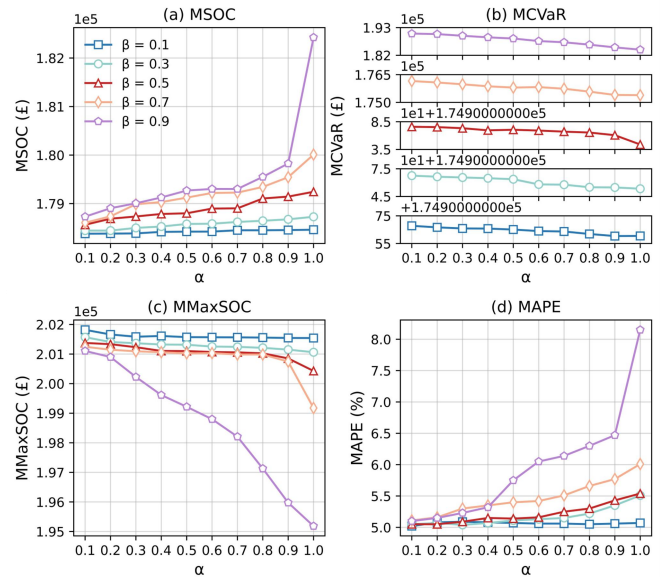

Fig. 6. Impact of varying parameters $\alpha$ on evaluation criteria ($\beta$ = 0.1, 0.3, 0.5, 0.7, 0.9).

$$\text{MAPE} = \frac{\hat{H} - H}{H} \times 100\% \quad (25)$$

*C. Basic Results*

Four forecasting models are used to forecast the system inertia from 2030/1/1 to 2030/1/7. The first three models are based on the RAOBF method. The parameters $\alpha$ of the three models are set to 0, 0.5, and 1, respectively, and the parameters $\beta$ are all 0.95, indicating the tail risk with a confidence level of 0.95. The last is an MSE-oriented forecasting model (MSEOF) based on the linear regression method [25].

The forecasting results of the four models are shown in Fig. 5(a). The RAOBF tends to under-forecast compared to the MSEOF. As the parameter $\alpha$ increases, the RAOBF will be more conservative. The regulating reserves calculated from the forecasted inertia values are shown in Fig. 5(b). More conservative forecasts will result in more regulating reserves being prepared in the day-ahead market.

Table III shows the performance of four forecasting models on MSOC, MCVaR, MMaxSOC, and MAPE. In terms of MAPE, the MSEOF has a better performance than the other three models because the objective of this forecasting model is the minimal statistical error. However, performing better on statistical criteria does not lead to better performance on economic criteria. Compared with MSEOF, the RAOBF ($\alpha$ = 0) reduces by 1.81%, 7.89%, and 7.65% on MSOC, MCVaR, and MMaxSOC, respectively. Although the RAOBF ($\alpha$ = 0) tends to under-forecast than MSEOF and results in more day-ahead costs, it avoids higher balancing costs and thus performs better on these three criteria. Compared with MSEOF, the RAOBF ($\alpha$ = 0.5) reduces by 1.24%, 9.19%, and 8.91% on MSOC, MCVaR, and MMaxSOC, respectively. The RAOBF ($\alpha$ = 0.5) is more conservative than the RAOBF ($\alpha$ = 0), resulting in a slight increase in MSOC but a decrease in MCVaR and MMaxSOC. Compared with MSEOF, the RAOBF ($\alpha = 1$) increases by 1.32% on MSOC but reduces by 9.48% and 11.2% on MCVaR and MMaxSOC, respectively. The RAOBF ($\alpha = 1$) is the most conservative forecasting, resulting in a significant increase in MSOC but a significant decrease in MCVaR and MMaxSOC.

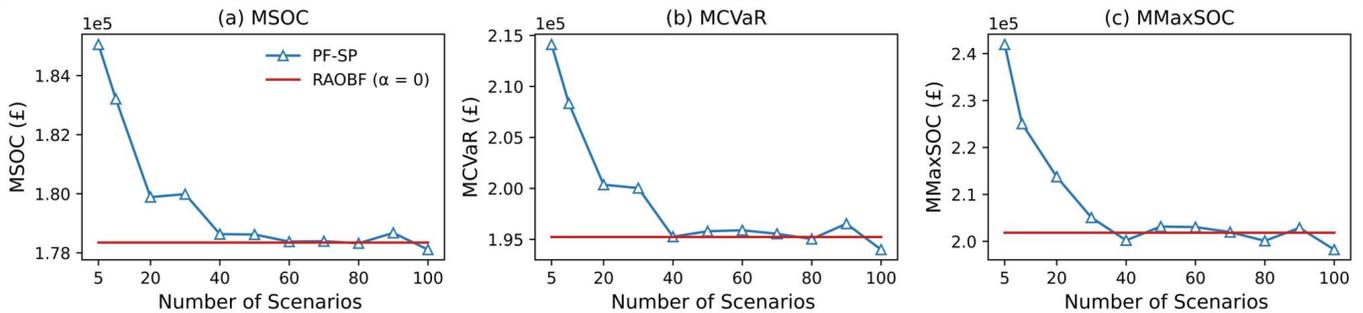

Fig. 7. Comparison of PF-SP and RAOBF ($\alpha = 0$) on MSOC, MCVaR, and MMaxSOC.

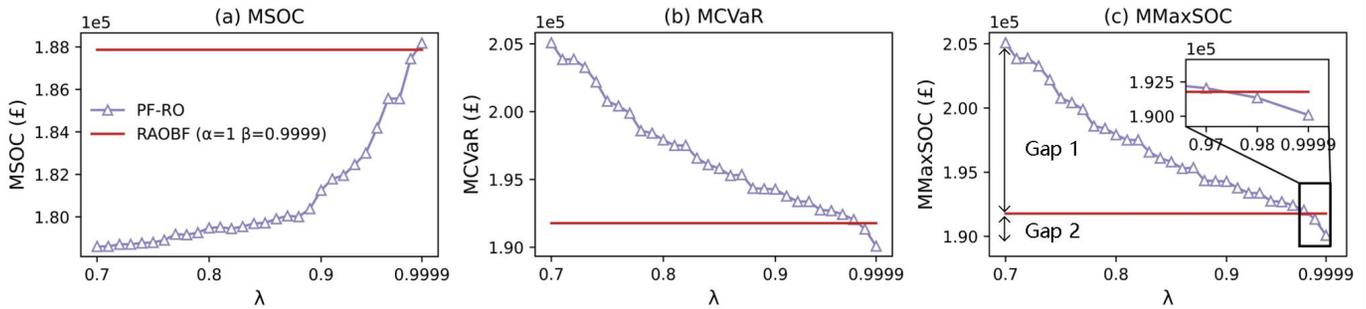

Fig. 8 Comparison of PF-RO and RAOBF ($\alpha = 0$ $\beta = 0.9999$) on MSOC, MCVaR, and MMaxSOC.

TABLE IV
COMPARISON WITH PF-SP WITH NUMBER OF SCENARIOS = 100

|  | MSOC (£) | MCVaR (£) | MMaxSOC (£) | Calculation time (s) |
|---|---|---|---|---|
| RAOBF ($\alpha=0$) | 178348.4 | 195235.0 | 201839.7 | 2.07 |
| PF-SP | 178108.7 | 193994.8 | 198243.5 | 165.43 |
| Gap (%) | **0.13** | **0.64** | **1.8** | **-7891.8** |

TABLE V
COMPARISON WITH PF-RO WITH $\lambda = 0.9999$

|  | MSOC (£) | MCVaR (£) | MMaxSOC (£) | Calculation time (s) |
|---|---|---|---|---|
| RAOBF ($\alpha=1, \beta=0.9999$) | 187866.7 | 191776.8 | 191776.8 | 2.07 |
| PF-RO | 188179.1 | 190084.2 | 190084.2 | 93.84 |
| Gap (%) | **-0.17** | **0.89** | **0.89** | **-4433.3** |

*D. Analysis of the Impact of Parameters α and β on Evaluation Criteria*

Five sets of experiments were conducted to analyze the impact of parameters $\alpha$ and $\beta$ on MSOC, MCVaR, MMaxSOC, and MAPE. The parameter $\alpha$ of each set of experiments is set from 0.1 to 1 with an interval of 0.1, and the parameter $\beta$ of the five sets of experiments is set to 0.1, 0.3, 0.5, 0.7, and 0.9, respectively. The results are shown in Fig. 6. The following points are observed:

1) Fixing $\beta$, the MPAE increases as $\alpha$ increases. Similarly, fixing $\alpha$, the MPAE increases as $\beta$ increases. Increasing $\beta$ and $\alpha$ will make the forecasts more conservative and thus increase the error between forecasted and true values.

2) Fixing $\beta$, the MSOC increases as $\alpha$ increases, while the MCVaR and MMaxSOC decrease as $\alpha$ increases. The increase of $\alpha$ means that impact of the risk part in the RAOBF method increases, leading to the model tending to reduce the risk of cost variability instead of the expectation of system operating costs. Therefore, the criteria MCVaR and MMaxSOC to measure the system operating risk will be improved, while the criteria MSOC to measure the economy will be worse.

3) Fixing $\alpha$, the MSOC increases as $\beta$ increases, while the MMaxSOC decreases as $\beta$ increases. This is because the closer $\beta$ is to 1, the more sensitive RAOBF is to tail risk, which is usually much higher than the average system operating cost. Therefore, to mitigate tail risks, it is often necessary to sacrifice more cost-effectiveness. Note that the criterion MCVaR cannot be compared here because this criterion is also a function of $\beta$.

These five sets of experiments illustrate that decision-makers can adjust the parameters $\alpha$ and $\beta$ according to their risk preferences to obtain satisfactory forecasting performance. For example, if the system operator wishes to mitigate the risk as much as possible, then the values of the parameters $\alpha$ and $\beta$ should be increased as much as possible. On the contrary, if the system operator wishes to have better cost-effectiveness, then the values of the parameters $\alpha$ and $\beta$ should be reduced as much as possible.

*E. Comparative Analysis with PF-SP and PF-RO*

*1) PF-SP vs RAOBF ($\alpha = 0$):* A two-stage SP model is used for comparison with RAOBF. The first stage of the model models day-ahead market clearing, as shown in (19), whereas the second stage models real-time market clearing, as shown in (20). To construct the scenario set $\mathcal{H}$, the quantile regression method is first used to construct a prediction interval with a confidence level $\lambda$ of 0.95, and then the scenarios are generated by Latin Hypercube sampling.

To illustrate the equivalence of RAOBF and PF-SP, we use the training scenario set $S$ to solve (21), the number of scenarios for PF-SP is set from 5 to 100, and the probability of each scenario is set to be the same. The performance comparison of PF-SP and RAOBF on MSOC, MCVaR, and MMaxSOC is shown in Fig. 7. Overall, the performance of PF-SP improves as the number of scenarios increases. When the number of scenarios is greater than or equal to 40, PF-SP and RAOBF have close performance on MSOC, MCVaR, and MMaxSOC, which means that using the training scenario set $S$ to solve (21) can also make the performance of RAOBF approximate that of PF-SP.

The comparison between RAOBF and PF-SP (number of scenarios = 100) is shown in Table IV. It can be observed that the performance of the two on MSOC, MCVaR ($\beta$ = 0.95), and MMaxSOC is very close, and the gap ((A-B)/B×100%) is less than 2%. The gap is due to the forecasting error and the difference between the training scenario set $S$ and the scenario set $\mathcal{H}$. RAOBF has a huge advantage in terms of computational cost compared with PF-SP. This is because the RAOBF method can be trained offline using historical data. When solving the deterministic model online, the system operator only needs to use a linear or nonlinear function to obtain the forecasted values required by the deterministic model. For PF-SP, constructing the scenario set is usually very time-consuming, and the computational cost of uncertainty optimization is much higher than that of deterministic optimization.

*2) PF-RO vs RAOBF ($\alpha$ = 1, $\beta$ = 0.9999):* A two-stage RO model is used for comparison with RAOBF, where the first and second stages model the day-ahead market clearing and the real-time market clearing, respectively.

To illustrate the equivalence of RAOBF and PF-RO, we use the training scenario set $S$ to solve (21), and the confidence level $\lambda$ of the box uncertainty set $\mathcal{U}$ is set from 0.7 to 0.9999 with an interval of 0.01 from 0.7 to 0.98. The performance comparison of PF-RO and RAOBF on MSOC, MCVaR, and MMaxSOC is shown in Fig. 8. The MSOC of PF-RO increases as $\lambda$ increases. RAOBF is close to PF-RO with $\lambda$ of 0.9999 on MSOC, which means that its cost-effectiveness is worse than PF-RO in most cases. However, on MCVaR and MMaxSOC, RAOBF performs better than PF-UO in most cases, which means it can better mitigate risks.

The comparison between RAOBF and PF-RO ($\lambda$ = 0.9999) is shown in Table V. It can be observed that the gap between RAOBF and PF-RO is also very small. There is only a -0.17% gap between the two on MSOC and 0.89% on MCVaR and MMaxSOC. The gap comes from the difference between CVaR and VaR, such as Gap1 in Fig. 8(c). When $\beta$ = 0.9999 and $\lambda$ = 0.9999, there is almost no difference between CVaR and VaR. However, there is still a gap between RACOF and PF-RO, such as Gap2 in Figure 8(c). The gap comes from the difference in the inertia forecasting between the probabilistic forecasting with a quantile of 0.0001 and the RAOBF with $\beta$ = 0.9999, as described in (12). Overall, when the parameters are matched, i.e., both $\lambda$ and $\beta$ are close to 1-, the performance of RAOBF can approximate that of PF-RO. RAOBF also has a huge advantage in terms of computational cost compared with PF-RO. The gap in performance and calculation time between RAOBF and PF-UO illustrates that RAOBF can approximate the performance of PF-UO at only a fraction of their computational cost.

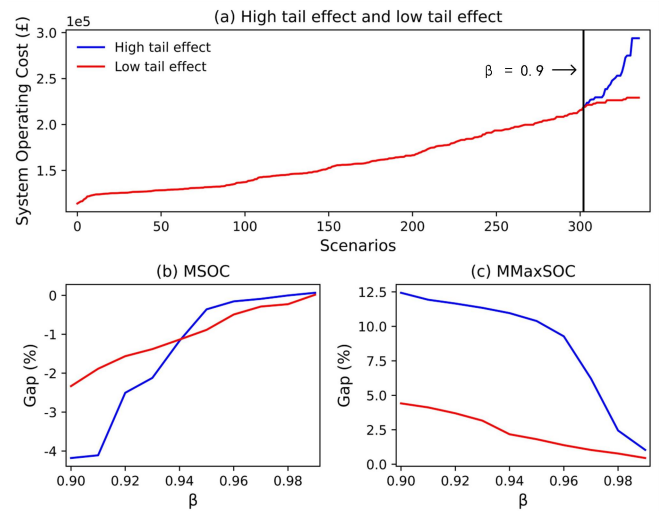

Fig. 9. Trend of the gap between RAOBF and PF-RO with $\beta$ for high and low tail effect cases.

We set up two cases to illustrate the effect of tail risk on the gap between RAOBF and PF-RO. These two cases have high and low tail effects, respectively. To illustrate this point, we arrange the system operating costs of each training scenario in the two cases in ascending order, as shown in Fig. 9(a). It can be observed that the variation range of the system operating cost after the quantile of 0.9 of the high tail effect case is significantly higher than that of the low tail effect case. As shown in Fig. 9(b), on MSOC, the same thing is that the gap decreases as $\beta$ increases in both the high and low tail effect cases. The difference is that MSOC is more sensitive to the change of $\beta$ in the high tail effect case. Similarly, as shown in Fig. 9(c), on MMaxSOC, the gap decreases as $\beta$ increases in both high and low tail effect cases, while MSOC is more sensitive to the change of $\beta$ in the high tail effect case. These two cases illustrate that the magnitude of the gap between RAOBF and PF-RO depends on the value of $\beta$ and the tail shape of the system operating cost distribution.

## VI. Conclusion

This paper proposes a RAOBF method that can effectively improve the effectiveness of its downstream decision-making while managing risk. The proposed method consists of an objective part and a risk part. A trade-off can be made between the risk part and the objective part by adjusting the parameters $\alpha$ and $\beta$. Through theoretical analysis and application in inertia management, we analyze that the proposed model enables the performance of deterministic models to approximate that of stochastic and robust models at only a fraction of the computational cost. Therefore, decision-makers can adjust the parameters $\alpha$ and $\beta$ according to their risk preferences to obtain satisfactory forecasts.